\documentclass[11pt,a4paper]{article}
\usepackage[hyperref]{aclrelex2020}
\usepackage{times}
\usepackage{amsmath}
\usepackage{graphicx}
\usepackage{xcolor}
\usepackage{latexsym}
\usepackage{tabularx}
\usepackage{multicol, blindtext}
\usepackage{float}
\usepackage{amssymb}
\usepackage{dblfloatfix}

\usepackage{microtype}
\usepackage{bbm}
\usepackage{subcaption} 
\usepackage{listings}
\aclfinalcopy 

\title{Relation Extraction with Explanation}

\author{Hamed Shahbazi, Xiaoli Z. Fern, Reza Ghaeini,  Prasad Tadepalli \\
School of Electrical Engineering and Computer Science, Oregon State University \\
Corvallis, OR, USA \\
\texttt{\{shahbazh, xfern, ghaeinim, tadepall\}@oregonstate.edu}
}

\date{}

\begin{document}
\maketitle
\begin{abstract}
Recent neural models for relation extraction with distant supervision alleviate the impact of irrelevant sentences in a bag by learning importance weights for the sentences. Efforts thus far have focused on improving extraction accuracy but little is known about their explainability. In this work we annotate a test set with ground-truth sentence-level explanations to evaluate the quality of explanations afforded by the relation extraction models. We demonstrate that replacing the entity mentions in the sentences with their fine-grained entity types not only enhances extraction accuracy but also improves explanation. We also propose to automatically generate ``distractor'' sentences to augment the bags and train the model to ignore the distractors. Evaluations on the widely used FB-NYT dataset show that our methods achieve new state-of-the-art accuracy while improving model explainability.\footnote{Accepted as a short paper at ACL 2020}

\end{abstract}
\section{Introduction}
Relation extraction with distant supervision associates a pair of entities with a bag of sentences, each containing mentions of both entities. The bag is tagged with relations between the pair in a Knowledge Base (KB), without explicitly indicating which sentence(s) support the relation(s). This method avoids the burden of manual annotations, but presents inherent ambiguity, creating challenges for learning. 

To alleviate the impact of the irrelevant 
sentences many approaches have been proposed including models based on attention \citep{l-zeng-15, l-lin-16, l-liu-17, l-luo-17, l-du-18, l-wang-18, l-missed-1-19, l-bai-19}, approaches that use additional resources \citep{l-vashishth-18, l-liu-18} and methods that utilize supervision data  \citep{l-pershina-14, l-angeli-14, l-iz-19}.
These studies primarily focus on improving relation extraction accuracy and little is known about whether the models are making right decision for the right reason or because of some irrelevant biases \citep{l-agrawal-16, l-gururangan-18, l-reza-19}. 

This paper examines two strong baseline relation extraction models with several explanation mechanisms. We manually annotated a test set from the widely used FB-NYT dataset with ground truth explanations to evaluate the quality of the explanation afforded by these models. We also introduce two different methods for improving relation extraction. First, we demonstrate that replacing the entity mentions with their fine-grained entity types for sentence representation leads to improvement in both the extract accuracy and model explainability. Second, we augment the bags with automatically generated ``distractor'' sentences (i.e., sentences that contain no supporting information for the relation) and train the model to appropriately ignore the irrelevant information. Our evaluation on the widely used FB-NYT dataset verifies that the proposed methods achieve the new state of the art for the extraction performance along with improved model explainability.
\section{Problem Setup}
Given entity pair $(e_i, e_j)$, we form a bag $B_{i,j}=\{s_1, \dots s_{N_{ij}}\}$ with $N_{ij}$ sentences that contain mentions of both entities and label it by the set of relations between $e_i$ and $e_j$ from the KB. Neural models for relation extraction encode each sentences into a vector representation and a bag $B_{i,j}$ is thus represented by $\{x_1, \dots x_{N_{ij}}\}$ where $x_i \in \mathbb{R}^{d}$. 

Given a set of bags and the associated labels, the training objective is to learn a model that predicts the probability $P(r=k|B_{i, j})$ that relation $k$ exists between $e_i$ and $e_j$ based on $B_{i,j}$, where $k\in 1\dots K$ and $K$ is the total number of relations in the KB. There are zero to multiple possible relation labels for each bag. Importantly, only some sentences in the bag express any of the relations and the others are irrelevant (provide no  information regarding the relations), but such sentences are not labeled.

\section{Baseline Models}
\label{sec:baselines}
We consider two baselines. The first is DirectSup, a recent model achieving the state-of-the-art performance by utilizing auxiliary supervision \cite{l-iz-19}. The second baseline (CNNs+ATT) revamps the classic attention based method by \citet{l-lin-16} but adopts the same sentence encoder as DirectSup for ease of comparisons. In this work, we add a ReLU at the end of the sentence encoder \cite{l-iz-19} to produce positive sentence representations. See \cite{l-iz-19} for detailed information regarding the sentence encoder. 

\noindent\textbf{DirectSup.} Given a bag of sentences, DirectSup encodes each sentence using CNNs with different filter sizes. The outputs of the CNNs with different filter sizes are concatenated to produce the encoding of the sentence. 

Given a bag $B$ and the encoding of its sentences $\{x_1, x_2,...,x_{N}\}$, DirectSup assigns an importance weight for each sentence based on the output of a binary classifier learned from an additional direct supervision data in a multi-task manner. Given a sentence encoding $x_n$, the binary classifier provides a weight $\alpha_n \in [0, 1]$ indicating the likelihood that $x_n$ expresses some form of relations in the KB. As a result, for a bag $B_{i,j}$, we have importance weights $\{\alpha_1, \dots, \alpha_N\}$. It then produces a single bag representation as follows:
\begin{equation}
    \bar{x} = \mbox{Max-pool}(\{\alpha_1 x_1, \dots, \alpha_n x_{N}\})
    \label{l:xrep_DirectSup}
\end{equation} 
and the prediction for relation $k$ is given by:
\begin{equation}
    P(r=k|B) = \sigma(\bar{x} \cdotp r_k + b_k)
    \label{lab:dir}
\end{equation}
where $r_k$ is an embedding of relation $k$, $b_k$ is a bias variable and $\sigma$ is the Sigmoid function.

\noindent\textbf{CNNs+ATT.} This model uses the same sentence encoder as DirectSup but differs in the attention mechanism used to decide sentence importance. Specifically, it follows \citet{l-lin-16} and computes the importance weights of the sentences in bag $B$ with encodings $\{x_1, \dots, x_N\}$ as follows:
\begin{equation}
    \alpha_{k,n} = \frac{\mbox{exp}(x_n A q_k)}{\sum_{i=1}^{N}\mbox{exp}(x_i A q_k)}
\end{equation}
where $q_k$ is a learned query vector associated with relation $k$ and $A$ is a diagonal matrix. 

Given $\{\alpha_{k,1}, ..., \alpha_{k,N}\}$, we compute a bag representation specific for relation $k$ by:
\begin{equation}
    \bar{x}_k = \sum_{n=1}^{N}{\alpha_{k,n} x_n}
    \label{l:xrepcnnsatt}
\end{equation}
and the prediction for relation $k$ is given by:
\begin{equation}
    P(r=k|B) = \sigma(\bar{x}_k \cdotp r_k + b_k)
    \label{lab:CNNatt}
\end{equation}
where $r_k$ is relation $k$'s embedding and $b_k$ is the bias.\\
\noindent\textbf{Entity embedding.} Prior work has demonstrated that incorporating entity embeddings into the relation extraction model leads to improved accuracy \citep{l-ji-17,l-iz-19}. Here we also consider this strategy with the baseline models. Specifically, let $v_i$ and $v_j$ be the entity embedding of $e_i$ and $e_j$, we concatenate the bag representations $\bar{x}$ with $v_i-v_j$ and $v_i\circ v_j$, where $\circ$ is element-wise product. We then apply a linear project layer with ReLU to produce a new bag representation for final prediction with Eq.~\ref{lab:dir} and \ref{lab:CNNatt}.

For any entity $e_i$ its embedding vector $v_i$ is obtained by concatenating the average of its skip-gram \citep{l-skip-13} word embeddings  and the embeddings produced by \citet{l-ernie-19} (produced by using TransE on Wikipedia factual tuples).

\noindent\textbf{Training objective.} For all the models in this work we use the binary cross entropy loss function for training:
\begin{equation}
\begin{aligned}
\label{l:loss}
    l = -\sum_{B_{i,j}}\sum_{k=1}^{K} \mathbbm{1}_{i,j,k} \hspace{1mm} \mbox{log} \hspace{1mm} P(r=k|B_{i,j}) +\\
    (1-\mathbbm{1}_{i,j,k}) \hspace{1mm} \mbox{log} \hspace{1mm} (1-P(r=k|B_{i,j}))
\end{aligned}
\end{equation}
where $\mathbbm{1}_{i,j,k}$ is an indicator function that takes value 1 if relation $k$ exists for bag $B_{i,j}$.

\section{Explanation Mechanisms}
\label{sec:measures}
The importance weights ($\alpha$'s, aka attention), generated by the models can be interpreted as explanations. However, recent studies \citep{l-reza-18,l-jain-19,l-sara-19} have questioned the validity of attention as a faithful explanation of model's behavior. Thus we consider the following additional explanation mechanisms:

\noindent\textbf{Saliency.} Recent works show that a model's prediction can be explained by examining the input saliency, based on the gradient of the output w.r.t. the inputs \citep{l-sal-1-13, l-sal-2-17, l-reza-19}. We define the saliency of sentence $n$ for relation $k$, denoted by $S_{x_n,k}$, as the L1 norm of the gradient of relation $k$ logit $o_k$ with respect to $x_n$.(Appendix.~\ref{sec:app_sal}).  

\noindent\textbf{Gradient $\times$ input.} This is a commonly used measure for input attributions \cite{l-shrikumar2016just,l-Selvaraju_2019}. We will refer to this measure as $GI_{x_n,k}$, computed as $\sum_i x_n[i] \times \frac{\partial o_k}{\partial x_n}[i]$. 

\noindent\textbf{Leave One Out (loo).} This measures the sensitivity of $o_k$ to the removal of a sentence.  We refer to this measure as $loo_{x_n, k} = (o_k-o_{k, -n})$, where $o_{k, -n}$ is the new logit of relation $k$ after removing sentence $x_n$ from its bag.  

\section{Proposed Methods}
\label{sec:model}
We propose two different approaches for improving relation extraction. The first method we propose, introduces a subtle change to the representation of the sentences, which lead to higher performance and better explanation quality. We further propose to automatically generate ``distractor'' sentences and train the model to appropriately ignore them.

\noindent\textbf{Sentence representation.} Each sentence in a bag contains entity mentions $m_i$ and $m_j$ for entities $e_i$ and $e_j$ respectively. In prior work $m_i$ and $m_j$ are kept unchanged \cite{l-lin-16,l-iz-19}. We argue that when entity mentions are used to compute the sentence representation, they provide such rich information that the model may not need to look at the rest of the sentence to deduce a relation. To ensure that our predictions are supported by appropriate sentences, we need to remove this effect. We propose to replace the entity mentions with their Fine-Grained Entity Types (FGET) \citet{l-figer-12} to force the model to identify the relations through the sentences. 

\noindent\textbf{Learning from distractors.} Prior work studied learning from human provided rationales \citep{l-missed-2-16, l-sal-2-17, l-rational-18, l-reza-19} in order to improve model explainability. However, human rationales are expensive to acquire. In this work we propose to learn from automatically generated ``distractor'' sentences. 

Let $B_{i, j}$ be a positive training bag (contains at least one relation) with entities $(e_i, e_j)$ of FGET $(t_i, t_j)$. Let $R_{ij} (|R_{ij}|> 1)$ be the set of annotated relations for $B_{i, j}$. For each $k$ in $R_{ij}$, we sample a ``distractor'' sentence $s'_k$ from the set of sentences in the training set such that 1) it belongs to a bag whose FGET is $(t_i, t_j)$ 2) the bag is not annotated with relation label $k$. If $s'_k$ is not found this way, we simply choose a random sentence from a random negative bag (bag with no relation). Given $s'_k$, we replace its entity mentions with $e_i$ and $e_j$ (or $t_i$ and $t_j$ for FGET-based sentence representation) of a sentence in $B_{i, j}$ and add it to the bag, resulting in an augmented bag $B'_{i,j}$ for relation $k$. 

To learn from the augmented bags, we feed $B'_{i,j}$ into the model and the goal is to lower the contribution of the distractor sentence in relation to the original sentences in the bag. Specifically, we use $GI$ to measure the sentence-level contribution and  define the distractor loss for relation $k$ as follows:
\begin{equation}
\begin{aligned}
l'_{d,k} = \mbox{max}(0, \gamma + GI_{x'_k, k} - \underset{x \in B_{i, j}}{\mbox{max}} GI_{x, k}) \\ + |GI_{x'_k, k}|
\label{eq:loss_noise}
\end{aligned}
\end{equation}
where $x'_k$ is the encoding of distractor sentence $s'_k$ and $\gamma$ is a hyper-parameter for margin. The first term ensures that the contribution of the distractor is lower than the maximum contribution of all the sentences in the original bag and the second term reduces the absolute contribution of the distractor. Although we use $GI$ in Eq.\ref{eq:loss_noise}, other explanation measures such as saliency or the positive portion of the contributions can also be applied here. Moreover a more advanced mechanism for generating distractors will likely lead to a higher performance.

We hence update the loss in Eq.~\ref{l:loss} with:
\begin{equation}
    l_m = l + \lambda l'_d
\end{equation}
where $l'_d=\sum_k l'_{d,k}$ and  $\lambda$ tradeoffs the regular learning loss with the distractor loss.

\begin{figure*}[!t]
\centering
\begin{minipage}[b]{.45\textwidth}
\includegraphics[width=\textwidth, height=6cm]{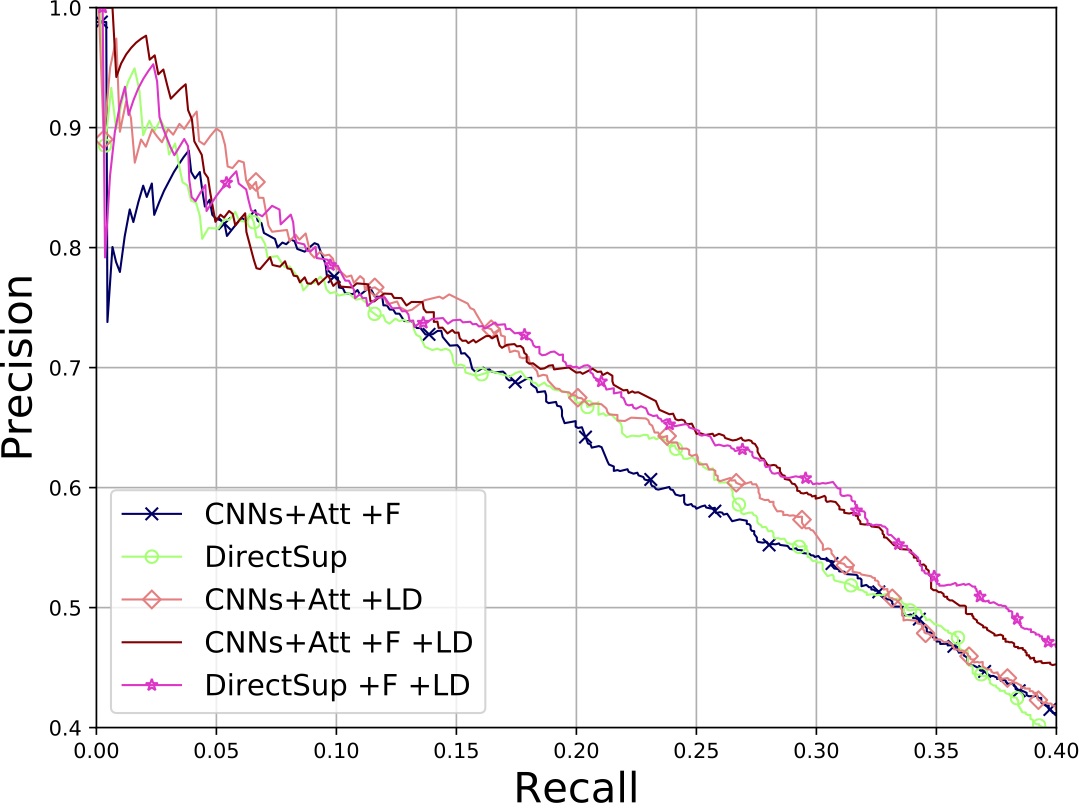}
\caption{PR without entity}\label{label-a}
\end{minipage}\qquad
\begin{minipage}[b]{.45\textwidth}
\includegraphics[width=\textwidth, height=6cm]{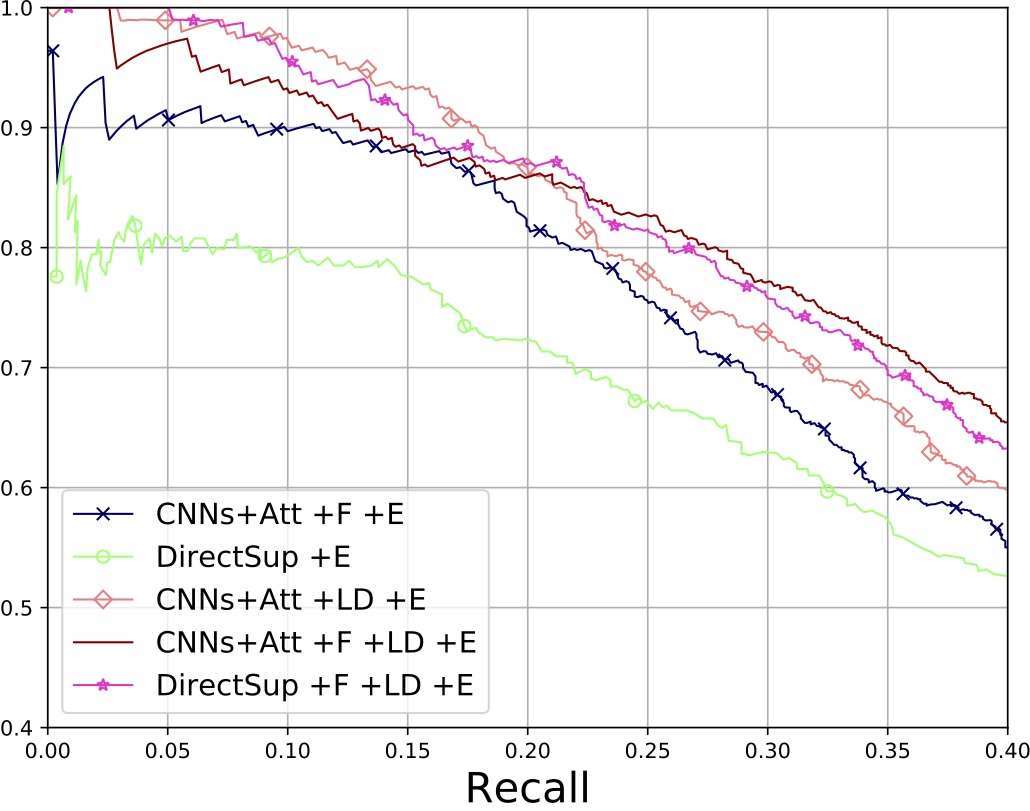}
\caption{PR with entity}\label{label-b}
\end{minipage}
\end{figure*}

\section{Experiments}
\label{sec:length}
In this section, we empirically evaluate our proposed methods both in terms of their relation extraction performance and their explainability.
\subsection{Dataset and Setup}
\textbf{Dataset.} Similar to our baselines and prior work, we use the modified version of FB-NYT dataset. The original FB-NYT dataset was built by \citet{l-reidel-10} on New York Times articles which was aligned to Freebase facts. It later was modified by \citet{l-lin-16}. There are 52 relations in this dataset where ``place lived'', ``captial'', ``neighborhood of'', ``natinality'' and ``location'' are the most frequent relations. Tab.~\ref{tab:1} shows the size of the modified dataset.
\begin{table}[h]
\centering
\small
\begin{tabular}{lcc}
\hline 
& \textbf{Train} & \textbf{Test} \\ \hline
Sentences &  472,963 & 172,448\\
Positive bags &  16,625 & 1,950 \\
Negative bags & 236,811 & 94,917 \\
\hline
\end{tabular}
\caption{\label{tab:1} FB-NYT modified dataset.}
\end{table}

\noindent\textbf{Setup and Training.} All models are implemented in PyTorch, trained with a Adam optimizer with learning rate 0.001 for a maximum of 30 epochs. We use  300-d skip-gram \citep{l-skip-13} word embeddings and FGET embeddings and 5-d position embedding. During training we freeze the word and entity embeddings. 
All reported results are averaged over three different random runs. 
We train on 90\% of the training set and keep the remaining 10\% for validation.
We select $\lambda$ from the set $\{0.01, 0.1, 1.0, 10.0, 100.0\}$ and set $\lambda=1.0$ based on validation AUC and the margin is fixed at $\gamma=0.00001$. 

\noindent\textbf{Ground-truth explanations.}
There are 1950 positive bags (6444 sentences) in the test split of FB-NYT. For each pair of sentence-relation in a bag we annotate whether the sentence entails the relation or not. Based on the annotations, we extract a set called \textit{expl-eval} (see Appendix~\ref{sec:app_gt} for details) including tuples of (bag-id, relation, positive sentence in bag, negative sentence in bag). Each tuple provides a desired ordering of two sentences when measuring their importance to the model. \textit{expl-eval} is then used to compute the Kendall Tau correlation between the annotation and the explanations, which measures how consistently the importance weights ranks the sentences compared to the ground truth.


\subsection{Relation Extraction Performance}

\begin{table}[t]
\centering
\small

\begin{tabular}{lcc}
\hline \textbf{model} & \textbf{AUC (-E)} & \textbf{AUC (+E)} \\ \hline 
\small CNNs+ATT  & 25.1 & -\\
\small DirectSup  & 26.4 & 28.1\\
\hline
\small CNNs+ATT +F & 26.1 & 31.5\\
\small DirectSup +F & 26.9 & 33.3 \\ 
\small CNNs+ATT  +FE & 27.4 & 33.1\\
\small DirectSup +FE & 27.6 & 33.4 \\
\small CNNs+ATT +LD & 27.1 & 33.6\\
\small CNNs+ATT +F +LD & 27.7 & 33.9\\
\small DirectSup +F +LD& \textbf{27.8} & \textbf{34.1}\\
\hline
\multicolumn{3}{l}{\tiny{F: Replace entity mention with FGET}}\\
\multicolumn{3}{l}{\tiny{FE: Replace entity mention with concatenation of FGET and entity mention}}\\
\multicolumn{3}{l}{\tiny{LD: Learning from distractor}}\\

\hline
\end{tabular}

\caption{\label{tab:auc} AUC results on FB-NYT.}
\end{table}

\begin{table*}[!t]
\centering
\small

\begin{tabular}{lcccccccc}
\hline \textbf{model} & loo (H) & loo (L) & $S_{x_{n,k}}$(H) & $S_{x_{n,k}}$(L) & $GI_{x_{n,k}}$(H) & $GI_{x_{n,k}}$(L)& $\alpha_{x_n}$(H) & $\alpha_{x_n}$(L) \\ \hline
\small CNNs+ATT & 0.16 & -0.08 & 0.19 & -0.02 &  0.20 & 0.04 & 0.69 & 0.21\\ 
\small DirectSup & 0.19 & 0.12 & 0.08 & 0.15 &  0.29 & 0.19 & 0.26 & -0.12\\\hline
\small CNNs+ATT +F & 0.21 & 0.10 & 0.36 & 0.03 &  0.23 & 0.00 & 0.73 & 0.11\\
\small DirectSup +F & 0.24 & 0.15 & 0.31 & -0.19 & 0.40 & -0.17 & 0.28 & 0.15 \\
\small CNNs+ATT +FE & 0.01 & -0.11 & 0.21 & -0.14 &  0.20 & -0.20 & 0.24 & 0.01\\
\small DirectSup +FE & 0.14 & -.12 & 0.19 & -0.10 & 0.29 & 0.06 & 0.17 & -0.11 \\
\small CNNs+ATT +LD & 0.18 & -0.01 & 0.22 & 0.10 &  0.21 & 0 & 0.67& 0.11\\
\small CNNs+ATT +LD +F & 0.22 & -0.11 & 0.43 & 0.09 & 0.28 & 0.07 & 0.70 & 0.12\\
\small DirectSup +LD +F & 0.23 & 0.14 & 0.38 & 0.01 & 0.49 & 0.20 & 0.45 & 0.02\\
\hline
\multicolumn{8}{l}{\tiny{H: High confidence $P(r) \in [0.76,1.0]$}}\\
\multicolumn{8}{l}{\tiny{L: Low confidence $P(r) \in [0,0.25]$}}\\

\end{tabular}

\caption{\label{tab:kendall} Kendall correlations for top confidence and least confidence range.}
\end{table*}

Similar to prior work we use precision-recall (PR)
curves to characterize the extraction performance and report the area under the PR curve (AUC) up to 0.4 recall. Tab.~\ref{tab:auc} reports the AUCs of the baselines and different variants of our proposed models with (+E) and without (-E) incorporating entity embeddings.

Specifically, we consider two different ways of incorporating the FGET representations. Rows 3-4 show the AUCs of the two baseline models when we replace entity mentions with their FGET (+F), whereas rows 5-6 show the AUCs when we concatenate the FGET with the entity mentions (+FE). From the results we can see that both baselines see clear performance gain from incorporating FGET into the representations.  Combining FGET with entity mention (+FE) achieves higher performance than using only FGET (+F), but our hypothesis is that the former will lead to less explainable models, which we will examine in the next section. 
Finally the last three rows of the table show that adding LD to different base models can further improve the AUCs. 

Similar to prior work, we observe that incorporating entity embeddings(+E) to the model leads to substantial performance gain across the board. We also observe very similar performance gain when adding FGET and LD to the base models both with and without entity embeddings. Our best model achieved an AUC of 0.341, which improves the previous state-of-the-art by 5.7\%.


\subsection{Evaluation of Explanations}
We apply the explanation mechanisms described in Section~\ref{sec:measures} to produce sentence importance scores for the test set and compute the Kendall Tau correlations for the importance scores using  \textit{expl-eval}.  

For each model, to understand its behavior when it predicts correctly versus incorrectly, we consider the subset $H$ ($L$) of bags/relations that the model outputs high (low) probability, i.e., $p\in [0.76,1]$ ($[0,0.25]$), for the correct relation. We report the performance on $H$ and $L$ separately in Tab.~\ref{tab:kendall}.

Comparing correlation values for $H$ and $L$ in Tab.~\ref{tab:kendall}, we observe that when the models are making correct and confident predictions ($H$), the values of correlation tend to be higher. In contrast, when the model fails to detect the correct relation ($L$), we see substantially lower correlation scores. 


By replacing entity mentions with their FGET in both CNNs+ATT and DirectSup (+F), we observe substantially increased correlation scores for correct predictions (H). The improvement is consistent across all methods that are used to compute the importance scores.  

Recall that Tab.~\ref{tab:auc} shows that concatenating FGET with entity mention (+FE) yields improved relation extraction performance for both CNNs+ATT and DirectSup. In contrast, the explanation results presented here show that this comes at the cost of explainability, as demonstrated by the substantially lower correlation scores of CNNs+ATT+FE and DirectSup+FE. This confirms our conjecture that removing entity mentions from the sentence representation leads to more explainable models, possibly by forcing the model to focus on the textual evidence contained in the sentence rather than the word embedding of the mentions.

Finally, we note that adding LD further improves the correlation score on $H$ for $S$, $GI$ and $\alpha$. This suggests that learning from distractors is a valuable strategy that not only produces better relation extraction performance, but also 
enhances the model explanability. 

\section{Conclusion}
In this work we provided an annotated test set with ground-truth sentence-level explanations to evaluate the  explanation quality of relation extraction models with distant supervision. Our examination of two baselines show that a model with lower relation extraction accuracy could have higher explanation quality. We proposed methods to improve both the accuracy and explainability. Our proposed methods are based on changing the representation of the sentences and learning from distractor to teach the model to ignore irrelevant information in a bag. Our evaluation on the widely used FB-NYT dataset show the effectiveness of our method in achieving state-of-the art performance in both accuracy and explanation quality.



\bibliography{aclrelex2020}
\bibliographystyle{acl_natbib}

\clearpage
\appendix

\section{Supplemental Material}
\label{sec:supplemental}

\subsection{Saliency and (Gradient $\times$ input)}
\label{sec:app_sal}
Assume that a neural model outputs a logit score $o$ which is a differentiable function and parameterized by $x \in \mathbb{R}^{d}$, $\theta$ and etc. The Taylor series of the given function $o$ near input $a$ is given by:
\begin{equation}
    o(x) = o(a) + \frac{\partial o}{\partial x}(a)(x-a) + \frac{1}{2!} \frac{{\partial o}^{2}}{\partial x^{2}}(a)(x-a)^{2} + \dots
\end{equation}
Approximating the function $o$ as a linear function, the first order approximation of the Taylor series is given by:
\begin{equation}
    o(x) \approx \frac{\partial o}{\partial x}(a)x + b
\end{equation}
Note that $\frac{\partial o}{\partial x}(a) \in \mathbb{R}^{d}$. Therefore for each dimension $i$ the bigger $\frac{\partial o}{\partial x}(a)[i]$ , the more (positive or negative) the impact of $a[i]$ is on $o$. The whole impact of $a$ on $o$ is given by $\sum_{i}\frac{\partial o}{\partial x}(a)[i]$ of its absolute value $\sum_{i}|\frac{\partial o}{\partial x}(a)[i]|$.

Regarding our task, the logit score of the model for a relation $k$ is $o_k$. For a given sentence $x_n$, the amount of positive or negative impact of $x_n$ on $o_k$ is approximated by $\sum_{i}|\frac{\partial o_k}{\partial x}(x_n)[i]|$ which is saliency.\\
The (Gradient $\times$ input) for a given sentence $x_n$ is equivalent to the linear approximation of $o_k$ at $x_n$ which is $\sum_{i} x_n[i] \times \frac{\partial o_k}{\partial x}(x_n)[i]$.

\subsection{Ground-truth explanation set.}
\label{sec:app_gt}
We annotate the positive bags of the test split of FB-NYT with ground-truth explanations. There are 1950 bags and 6444 sentences. For each pair of (sentence, relation) in a bag, the sentence is either a rationale (supportive) to the relation or it is irrelevant. For example:

\begin{table}[h]
\scriptsize
\centering
\begin{tabularx}{.48\textwidth}{X}
\hline
entity pair: (\colorbox{gray!30}{namibia}, \colorbox{gray!30}{windhoek})\\
relation: /location/country/capital\\
\hline
\colorbox{green!30}{rationale}:
```the magistrate also continued mr. alexander 's bail conditions 
, including a bond of 10 million namibian dollars about  1.4 million 
and restrictions on his movements to the magisterial district 
of \colorbox{gray!30}{windhoek} , \colorbox{gray!30}{namibia} 's capital```\\

\colorbox{red!30}{irrelevant}:
```mr. alexander also placed full page ads in local newspapers proclaiming 
his commitment to investing in \colorbox{gray!30}{namibia} , 
and has mounted a large billboard conveying the same message 
opposite government park in \colorbox{gray!30}{windhoek}```\\
\hline
\end{tabularx}
\end{table}

Following the annotation of the sentence-relation contributions which is either rationale or irrelevant, we extract a set ``expl-eval'' (which is going to be used to evaluate the explanation quality of the models) as follows: 

\scriptsize
\begin{lstlisting}[escapeinside={(*}{*)}]
expl-eval = set()
For each (bag-id, bag):
 For each relation label (*$k$*) given to the bag:
    For each pair of rationale (*$s^{+}$*) an irrelevant (*$s^{-}$*)
       for (*$k$*):
            expl-eval.add((bag-id, k, (*$s^{+}$*), (*$s^{-}$*)))
     
\end{lstlisting}

\normalsize
The size of the generated expl-eval is 1097 tuples of (bag-id, $k$, rationale sentence, irrelevant sentence). Please note that the relation label $k$ is one of the ground-truth labels assigned to bag-id.

\end{document}